\documentclass{article}

\usepackage{arxiv}

\usepackage[utf8]{inputenc} 
\usepackage[T1]{fontenc}    
\usepackage{hyperref}       
\usepackage{url}            
\usepackage{booktabs}       
\usepackage{amsfonts}       
\usepackage{nicefrac}       
\usepackage{microtype}      
\usepackage{lipsum}
\usepackage{authblk}
\usepackage[inline]{enumitem} 
\usepackage{tasks}
\usepackage{amssymb}
\usepackage{algorithm}
\usepackage{tikz}

\title{modeling food popularity dependencies using social media data}

\author{\textbf{Devashish Khulbe} $\hspace{0.1cm}$ \& $\hspace{0.1cm}$ \textbf{Manu Pathak}\\Center for Urban Science + Progress\\ New York University\\ New York, NY 10003\\ {\{dk3596, mp4515\}@nyu.edu}}

\begin{document}
\maketitle

\begin{abstract}
The rise in popularity of major social media platforms have enabled people to share photos and textual information about their daily life. One of the popular topics about which information is shared is food. Since a lot of media about food are attributed to particular locations and restaurants, information like spatio-temporal popularity of various cuisines can be analysed. Tracking the popularity of food types and retail locations across space and time can also be useful for business owners and restaurant investors. In this work, we present an approach using off-the shelf machine learning techniques to identify trends and popularity of cuisine types in an area using geo-tagged data from social media, Google images and Yelp. After adjusting for time, we use the Kernel Density Estimation to get hot spots across the location and model the dependencies among food cuisines popularity using Bayesian Networks. We consider the Manhattan borough of New York City as the location for our analyses but the approach can be used for any area with social media data and information about retail businesses.
\end{abstract}


\section{Introduction}
In the past decade, numerous social media platforms like Instagram, Facebook and Twitter have seen a surge in popularity among users \cite{socialMedia} and generates vast amount of data in the form of photos, videos and text. Out of the many topics and categories that can be attributed to this data, one of the top ones is regarding food. A majority of posts on social media platforms about food are attributed to many retail businesses and restaurants. We argue that data referenced to defined locations can inform about the activity and potential hot spots regarding a particular food type or cuisine. Analyzing the time and number of posts about these locations can give insights about popularity among the online populace. Additionally, it can give an idea about what cuisines are popular in which area if we further dive deeper into analyzing the content of the food related media being posted. With the influx of information regarding food online, business owners can identify emerging areas of cuisine popularity and accordingly plan their future decisions to maximize their profits. Also, restaurant investors can direct their resources in a much more informed and dynamic way instead on relying on traditional census data based approaches for deciding to invest in a new location or business.

\subsection{Related work}
Data about social media platforms like Instagram and Twitter have been analyzed, indicating that posts about food forms major percentage of all media. Literature has also emphasized the potential impact of 'Instagram Branding' \cite{instaBranding} for the success of a business. Another research harvests data from social media platforms to identify momentary social hot spots based on sudden or gradual increase in online activity by users \cite{imp1}, emphasizing our hypothesis that online activity can inform about the physical activity for an area. Regarding analysis of users' posts about restaurants, there have been some work on sentiment analysis \cite{sentAnalysis} of content indicating that there is a scope of identification of areas with decreasing popularity along with emerging areas in our work. Furthermore, social media images have also been used as data for prediction of human characteristics and sentiment. Also, deep learning frameworks like Keras and PyTorch have been used to detect food cuisines from images with significant accuracy. 

\section{Data}
\label{sec:headings}

\subsection{Yelp}
Yelp API allows to extract information like name, average rating, cuisine type for restaurants in a location. We extract a total of 1000 operating (as of May 2019) restaurants for Manhattan. According to Yelp, the restaurants are randomly sampled out of the total number of restaurants. We then convert the name of each restaurant into corresponding hashtags, attributed for which we extract data from the social media. For instance, if a restaurant has name "Sample Name", the corresponding hashtag would be "\#SampleName". The rationale behind converting to hashtags is that social media users often write the hashtag name along their posts. This enable us to extract data regarding a particular topic, which in this case are food types.\\
The documentation for Yelp API may be found at
\begin{center}
  \url{https://www.yelp.com/developers/documentation/v3}
\end{center}

\subsection{Social media images}
Most recent images were collected corresponding to each location's (restaurant) hashtag from social media (mainly Instagram). A limit of 300 was put for each location's maximum media extraction. These were the most recent images posted by users corresponding to each hashtag. We anticipate that collecting 300 images would cover at least a 15 days worth of media for each restaurant. It was noticed that the posts extracted were dated to more than a year ago for some locations. We discarded these posts from our data and considered media for the last one year (since May 2018). Furthermore, the locations for which no media was found were also removed from the data set.

\subsection{Google Images}
The images for different cuisines and food types were taken from Google. This was required to fine tune a food classification model which we deployed for identifying different food types in social media data and also for removing non-food images from it. A total of 100 images were taken for each food type in a cuisine. This allowed for transfer learning for the classification model based on the images.
The 8 food types we looked here are:
\subsubsection{Food categories}

\begin{tasks}[style=itemize, column-sep=-35mm, label-align=left, label-offset={0mm}, label-width={3mm}, item-indent={12mm}](4)%
\task Ramen
\task Sushi
\task Waffles
\task Burgers
\task Hot Wings
\task Nachos
\task Bagels
\task Pizza
\end{tasks}

\section{Methods \& Results}
\label{sec:others}

\subsection{Image food classification model}
To classify the social media images into food classes, we used a pre-trained deep learning model and used transfer learning with our Google images.
The prediction model was a deep learning network (convolution neural net) trained using Keras. Transfer learning methodology was used on a pre-trained ResNet-50 model \cite{resnet} without the top layers. Training was switched off for the pre-trained layers since the base layers represent simple features applicable to most computer vision classification models. The base model was followed by blocks of fully connected layers and dropout layers for which training was switched on. The fully connected layers were sized based on the length of the feature vector returned by the base model. An Stochastic Gradient Descent (SGD) optimizer was used along with categorical cross entropy loss function. The model was trained for 200 epochs with a batch size of 32 per epoch. There were 800 total images for 8 types of food classes which were divided as 80:20 for training and validation. The test set were 200 manually annotated images from the social media data. The final accuracy of the model on the entire data set was 90\% with good performance across all 8 food classes.\\

The non-food images posted by users were thus discarded from the data set, which then resulted into a total of around 85,000 images. Furthermore, to adjust for time, we divided each day based on time as follows:

\subsubsection{Time slots}
\begin{itemize}
\item 5am-12noon : Breakfast
\item 12noon-6pm : Lunch
\item 6pm-2am : Dinner
\end{itemize}

\subsection{Kernel Density Estimation}
We used the Kernel Density Estimation (KDE) model using standard Gaussian kernel to get hot spots and interpolations for each 8 cuisines, optimizing for the bandwidth. The KDE function is defined as \\
\begin{equation}
\ \hat{f}_{h}(x)=\frac{1}{n}\sum _{i=1}^{N}K_{h}(x-x_{i})= \frac{1}{nh}\sum _{i=1}^{N}K(\frac{x-x_{i}}{h})
\end{equation}

where (x1, x2, …, xn) be a univariate independent and identically distributed sample drawn from some distribution with an unknown density f, K is the kernel function and h > 0 is the bandwidth.\\
The Gaussian kernel function is defined as: \\
\begin{equation}
\ K(u)=\frac{1}{\sqrt{2\pi}}e^-\frac{u^2}{2}
\end{equation}
Assuming that time of the day may be a confounder for number of posts and popularity of some food, we decided to do the analyses separately for three different times of the day.

The optimized bandwidth for the three slots is shown in Table 1. The KDE analysis gives an idea about the hot-spots in the location for different food types, adjusted for time. 
\begin{table}[h!]
  \begin{center}
    \caption{Optimized bandwidth parameters}
    \label{tab:table1}
    \begin{tabular}{l|c|r} 
      \textbf{Time Slot} & \textbf{KDE Bandwidth (in meters)} & \textbf{Hot spots}\\
      \hline
      5am-12noon & 134 & Lower East side, Chinatown\\
      12noon-6pm & 151 & Lower East side\\
      6pm-2am & 169 &  Hell's Kitchen, East Village\\
    \end{tabular}
  \end{center}
\end{table}

\subsection{Bayesian Network Modeling}

To model the dependencies among the popularity of food cuisines, we construct a Bayesian Network with BIC scoring method using the Greedy Hill Climbing approach for faster computations. Note that we only use our social media data (number of posts) to learn the structure of the network. Training and making predictions on the same data could be an area of further research. Essentially, the approach learned a Directed Acyclic Graph (DAG) \textit{G}(\textit{V}, \textit{E}) by modelling dependencies based on the number of posts of different food types in the last one year. To account for time of day, we also divided the data into three time slots as mentioned in section 3.1.1 and learned separate DAG structures for each slot.

\paragraph{Structure learning based on Maximum Likelihood Estimation and Hill Climb approach}
The search is started from either an empty, full, or possibly random network. Then given a Bayesian Network (BN) structure, the parameters of local probability distributions (pdfs) are estimated. In this case, these are the maximum-likelihood estimation of the probability entries from the data set, which for multinomial local pdfs consists of counting the number of tuples that fall into each table entry of each multinomial probability table in the BN. The algorithm’s main loop consists of attempting every possible single-edge addition, removal, or reversal, making the network that increases the score the most the current candidate, and iterating. The process stops when there is no single-edge change that increases the score.
The BIC score for DAG \textit{G} and data set \textit{D} is defined as:

\begin{equation}
\ BIC(G, D)=logP_r(D|\hat{p},G) - \frac{d}{2}logN
\end{equation}

The term $\hat{p}$ is the set of maximum-likelihood estimates of the parameters \textit{p} of the BN, \textit{d} is the number of dimensions of multivariate Gaussian and \textit{N} is the number of points in data.\cite{thesis}

The learned graph structure \textit{G(V, E)} for the three time slots are shown: \\

\textbf{G\textsubscript{5am-12noon}}: (Burgers$\rightarrow$ Nachos), (Burgers$\rightarrow$ Pizza), (Burgers $\rightarrow$ Bagels), (Burgers $\rightarrow$ Ramen), (Burgers $\rightarrow$ Sushi), (Burgers $\rightarrow$ Hot Wings), (Burgers $\rightarrow$ Waffles)

G\textsubscript{12noon-6pm}: (Ramen $\rightarrow$ Nachos), (Ramen $\rightarrow$ Pizza), (Ramen $\rightarrow$ Bagels), (Ramen $\rightarrow$ Burgers), (Ramen $\rightarrow$Sushi), (Ramen $\rightarrow$ Hot Wings), (Ramen $\rightarrow$ Waffles)

G\textsubscript{6pm-2am}: (Burgers $\rightarrow$ Nachos), (Nachos $\rightarrow$ Hot Wings), (Hot Wings $\rightarrow$ Ramen), (Ramen $\rightarrow$ Pizza), (Ramen $\rightarrow$ Sushi), (Pizza $\rightarrow$ Bagel), (Pizza $\rightarrow$ Waffles)

where each parent-child relationship is shown as (\textit{Parent} $\rightarrow$ \textit{Child})

\section{Discussion}
The Bayes Network structures for the three defined time slots models the dependencies among the 8 food types. Considering the number of posts as a proxy for popularity, we can infer some relationships among these food types in different locations. At the same time, it cannot be concluded that the resulted structures are the true causal dependencies for our data as we have not included variables like demographics and income which might be confounders in our analyses. That said, we aim to learn the Bayes Network structures at much more dynamically (update every month) than the change of other possible variable like demographics and median income, the change of which takes place in a time frame of years. Thus, we can assume that these confounding variables would not have much impact on the learning from our dynamically updated geo-tagged posts data. We adjust for time in this work by doing separate analyses for three time slots in a day. One can argue that most people are active on social media during the evening hours and thus there would be more number of posts during that time slot, making the data biased. We believe this is one of the limitations in our data and controlling for it could be an area of future work. Furthermore, the KDE gives the locations of hot spots which are indicative of the trending areas in our considered location. The results demonstrate that areas like Midtown and East Village are among the popular areas in food. The optimized bandwidth parameter is indicative of the radius within which nearby restaurants are influencing the popularity of an area. Through KDE, we also expect to know when a new area is emerging as a food hub or is becoming popular among people. Apart from KDE, we can also check spatial autocorrelation to see the spatial dependencies among the food types through measures like Moran's I and Anselin Local Moran's I in further work.

\section{Conclusion}
In this work, we postulate that popularity of businesses and food types in a location can be analyzed using the social media activity of users. We then present ways using machine learning techniques to identify the popular locations and model the dependencies (of popularity) among different food types in Manhattan, New York. The key areas of future work regarding this work include incorporation of other covariates like demographics and median income. Also, finding a proper metric to validate the results and predictions from these methods could be important to test our hypothesis. We measured the activity of users through social media images but also talked about including other media such as text. In addition to texts, information like business check-ins can also be used to expand this work and get comprehensive results. We believe that our approach can be important to business owners and investors who can make decisions based on the dynamically changing online behavior and opinions of users. This would be improvement from the current practice of looking at just the demographics data. Considering the short life cycle of restaurants, especially in New York City, we believe that incorporating new techniques and dynamic data is important and should be widely adopted in this industry.

\bibliographystyle{unsrt}  


\end{document}